\documentclass[12pt]{article}
\usepackage[koi8-ru]{inputenc}
\usepackage[english]{babel}
\usepackage{epsf}
\usepackage{pazh}
\usepackage{amsmath}	
\usepackage{amssymb}	

\newcommand{\ergs}{\,erg\,s$^{-1}$}
\newcommand{\kms}{\,km\,s$^{-1}$}

\newcommand{\gcmq}{\,g\,cm$^{-3}$}
\newcommand{\cmsqg}{\,cm$^2$\,g$^{-1}$}
\newcommand{\cmq}{\,cm$^{-3}$}
\newcommand{\gcm}{\,g\,cm$^{-1}$}
\newcommand{\ha}{H$\alpha$}

\begin{document}

\title{\bf Early bolometric luminosity of SN~2013fs without photometry  
}
\author{\bf \hspace{-1.3cm}\copyright\, 2020  \\
N. N. Chugai
}
\affil{
Institute of Astronomy of Russian Academy of Sciences, Moscow\\ 
}

\vspace{2mm}

\sloppypar 
\vspace{2mm}
\noindent


\noindent

\vfill
\noindent\rule{8cm}{1pt}\\
{$^*$ email $<$nchugai@inasan.ru$>$}

\clearpage

\centerline{\bf Abstract}
\vspace{0.5cm}

The novel method is proposed for the reconstruction of the early bolometric light curve 
  for supernovae IIP based on a set of spectra of \ha\ emission during the first day 
  after the shock breakout.  
The method exploites an effect of the radiative acceleration of the preshock circumstellar gas 
  that is manifested in the broad \ha\ wings.
The efficiency of the method is demonstrated in the case of SN~2013fs with 
 spectra taken between 6 and 10 hours after the shock breakout.
The exceptional feature of this method is that it does not require 
 the photometry, the distance, and the  extinction.
  
\vspace{0.5cm}

\section{Introduction}

A type IIP supernova (SN~IIP) is an outcome of the explosion of 
 a red supergiant (RSG) caused by the core-collapse. 
Following the explosion, ejecta expands in the environment 
  created by the presupernova mass loss.
The RSG wind of pre-SN~IIP  has a  moderate density parameter 
  $\dot{M}/u \sim 10^{14}-10^{15}$\gcm\ indicated by the radio and X-ray emission 
    (Chevalier et al. 2006). 
The optical emission lines, e.g., \ha\ from the wind of that density are generally expected to be weak 
  for the detection.
Yet in several SNe~IIP the spectra show during first 
  couple days rather strong narrow emission lines apparently originated from a confined 
   ionized dense circumstellar (CS) shell 
  (Quimby et al. 2007; Groh et al. 2014; Khazov et al. 2016; Yaron et al. 2017).
The most interesting case  is SN~2013fs with spectra starting 
  from 6\,h after the shock breakout (Yaron et al. 2017).
The analysis of these spectra led authors 
   to conclude that the early CS emission lines originate from the confined shell  
   ($r < 10^{15}$\,cm) with the mass of (several)$\times10^{-3}$\msun,
    the expansion velocity of 50-100\kms, and the Thomson optical depth of 
    $\tau_{\sc T} \sim 1-2$ (Yaron et al. 2017).
    
 The line profiles of early spectra of SN~2013fs show narrow core and extended wings 
  very much similar to emission lines of early SN~1998S, in which case they 
   have been attributed to the  emission from the dense CS shell with the optical depth to 
   the Thomson scattering  of $\tau_{\sc T} \sim 3-4$ (Chugai 2001).
However this interpretation fails for SN~2013fs because broad wings observed in 
   \ha\ are more intense compared to the model of the Thomson scattering; moreover the 
   blue wing is stronger than the red one unlike the model (Yaron et al. 2017).  
The problem has been resolved in the model that includes the acceleration of the  
  preshock CS gas by the SN radiation (Chugai 2020a); for the \ha\ at 10.3\,h the recovered 
  preshock CS velocity is 3000\kms. 
 The broad \ha\ wings in early spectra of SN~2013fs are essentially dominated by the 
  emission from the  accelerated CS gas and not by the Thomson scattering. 
 Remarkably, in SN~1998S the effect of the radiative acceleration of the circumstellar gas 
   upto 1000\kms\ is overwehlmed by the Thomson scattering (Chugai 2001).
Dessart et al. (2017) explored numerically effects of dense CS shell around SN~IIP using 
  radiation hydrodynamics and non-LTE radiation transfer. 
The models predict significant acceleration of the CS gas upto $> 5000$\kms\ during 
  first couple days.
However the reported synthetic spectra do not indicate, in which cases line wings 
  are dominated by either Thomson scattering or the bulk velocity of the accelerated CS gas.  

The proposed explanation  of the the early \ha\ spectrum of SN~2013fs prompts a   
  possibility that the preshock velocity  
  inferred from the early \ha\ can be used to probe the early bolometric luminosity of SN~2013fs 
  that otherwise is poorly determined.  
Here I present the method for the 
  reconstruction of the early bolometric light curve of SN~2013fs based on  
  preshock velocity extracted from the \ha\ emission line. 
To this end we first determine velocities of the preshock CS gas from the \ha\ 
  in Keck-I spectra between 6\,h and 10\,h after the shock breakout.
The found velocities are used then to recover the bolometric luminosity 
  of SN~2013fs during first 10 hours  after the shock breakout.
The study exploits spectra of SN~2013fs retrieved from the database  
  WISeREP (Yaron \& Gal-Yam 2012) ({\em https://wiserep.weizmann.ac.il}).

\section{Model}

The shock breakout of an exploding star with an extended envelope is accompanied with the 
  sweeping of the outmost  atmosphere into the dense shell (Grasberg et al. 1971).
For the exploding RSG the mass of the thin dense shell is 
 $10^{-4}-10^{-3}$\msun\  (Chevalier 1981).
The thin shell decelerates in the CS gas that results in the formation of 
 the forward and reverse  shock with the thin dense shell inbetween (Chevalier 1982a; Nadyozhin 1985). 
The reverse shock is generally radiative,   
  so the shocked cold ejecta are accumulated in the thin dense shell that we dub the 
   cold dense shell (CDS) since its kinetic temperature is lower than the temperature of 
   both shocks.   
   
 The size and density of the dense CS shell of SN~2013fs can be 
   recovered from spectral data in the following way.
 On day 2.42 the spectrum shows the broad He\,II 4686\,\AA\ emission (Bullivant et al. 2018) 
  that is attributed to the fragmented CDS with the expansion velocity of 
  $v_{cds} = 16600$\kms\ (Chugai 2020b).
Narrow \ha\ associated with the CS shell disappeares between epochs of Keck-II 
 spectra on day 2.1 and 5.1 (Yaron et al. 2017).
This means that at about $t \sim 3$\,d the CS shell is overtaken by the CDS,
  which gives us the extension of the CS shell $R_{cds} \sim v_{cds}t \sim 5\times10^{14}$\,cm.
The \ha\ modelling in the spectrum on 10.3\,h implies the Thomson optical depth 
  of the CS shell $\tau_{\sc T} \sim 2$ (Chugai 2020a).
The average electron number density in the CS shell is thus  
  $n_e = \tau_{\sc T}/(R_{cs}\sigma_{\sc T}) \sim 6\times10^9$\cmq\ and the 
   density $\rho_0 = 1.2\times10^{-14}$\gcmq\ assuming hydrogen abundance $X = 0.7$.
Below we adopt homogeneous CS shell with the density $\rho_0 = 1.35\times10^{-14}$\gcmq\ that 
  is by a factor of three higher compared to the model that is aimed at the minimization of 
  the explosion energy of SN~2013fs (Chugai 2020b).

\begin{figure}
    \epsfxsize=19cm
	\hspace{-2cm}\epsffile{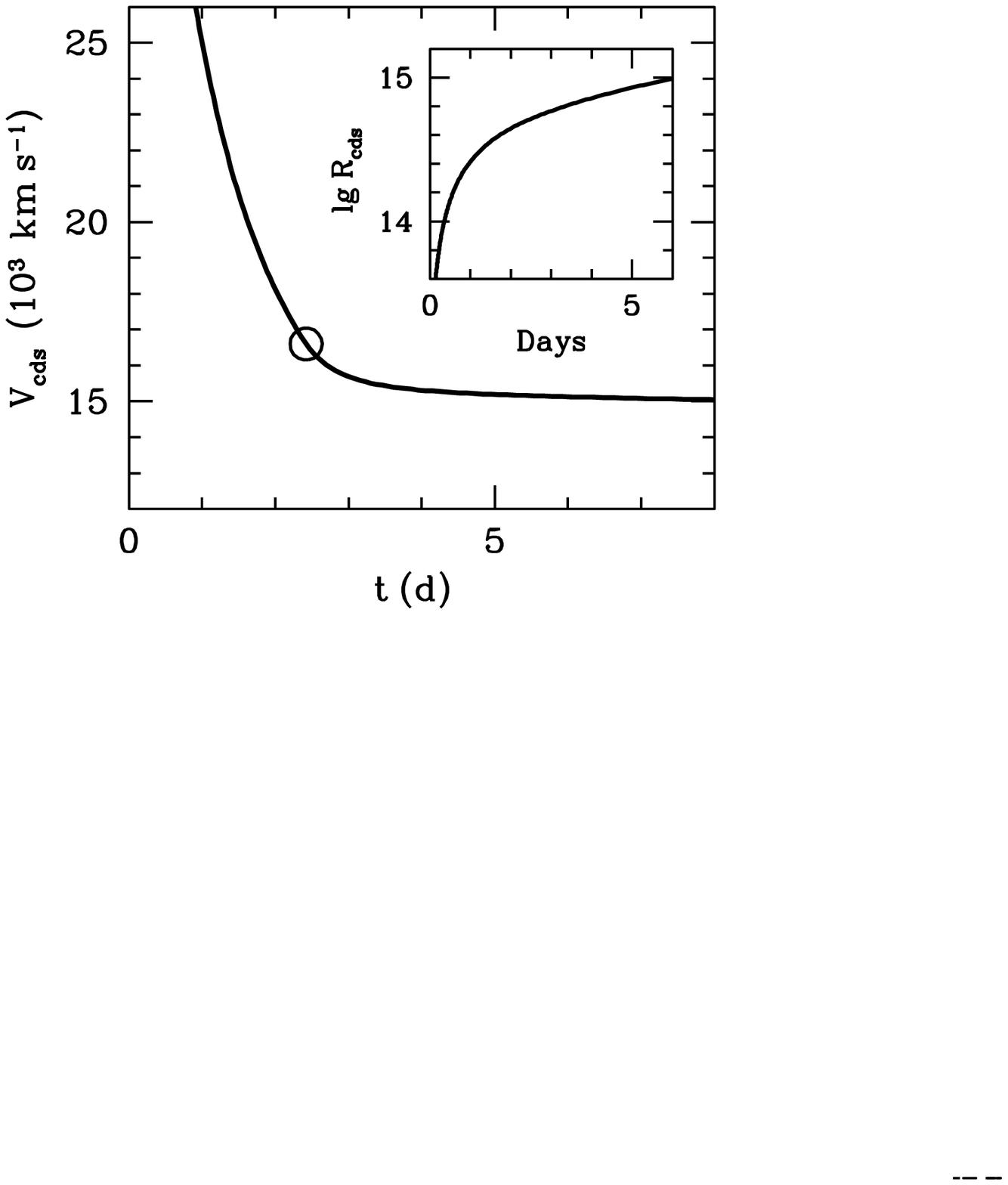}
	\caption{
	The model velocity of the cold dense shell fits the obeservational estimate 
	({\em circle}) from He\,II 4686\,\AA\ line on day 2.4. {\em Insert} shows the 
	evolution of the model CDS radius.
	    }
	\label{fig:dyn} 
\end{figure}

One can apply the hydrodynamics in the thin shell approximation 
  (cf. Chugai 2020b) to find the CDS dynamics that matches the expansion velocity on day 2.42.
The rate of the CDS decceleration is determined by the CS density and the SN ejecta 
   density distribution in outer layers; the latter generally follows the power law 
   $\rho(v) = \rho_1(t_1/v)^3 (v_1/v)^q$.
We adopt $q = 7.6$ in line with the hydrodynamic model of the normal type IIP 
 SN~2008in (Utrobin \& Chugai 2013).  
For the reference values $t_1 = 1$\,d, $v_1 = 10^4$\kms\ the CDS expansion 
  satisfies the velocity constrain on day 2.42 (Fig. \ref{fig:dyn}) for 
  $\rho_1(t_1,v_1) = 3.44\times10^{-10}$\gcmq.
  
At the initial phase of $\sim 1$ day the photosphere of SN~2013fs coincides with the CDS.
Indeed, the momentum flux conservation suggests that the CDS density 
 is $\sim \rho_0(v_{cds}/c_s)^2 \sim 6\times10^{-9}$\gcmq\ (Grasberg et al. 1971),
  where $c_s \approx 37$\kms\ is the isothermal sound speed for the CDS temperature of
  $T \sim (L/4\pi r^2\sigma)^{0.25} \sim 3.5\times10^4 L^{0.25}_{43}/r_{14}^{0.5}$\,K. 
For these conditions the Rosseland opacity $k_{\sc R} \sim 2$  (Badnell et al. 2005).
With the CDS mass of $\sim 3\times10^{-4}$\msun\ and the CDS radius at $t = 10$\,h 
   of $R_{cds} \sim 10^{14}$\,cm (Fig. \ref{fig:dyn} ) the CDS optical depth turns out 
   to be $\tau = k_{\sc R}M_{cds}/(4\pi R_{cds}^2) \sim 10$, so indeed the CDS at the 
   considered phase is opaque and the photosphere resides at the CDS.
   
The overall picture for the \ha\ formation at the considered stage thus 
  can be described as a photosphere (CDS) of the radius  $R_{cds}$ enclosed by the hot 
  forward shock of 
   the radius $R_s \approx \xi R_{cds}$ that is embedded into the ionized confined CS shell 
   $R_s < r < R_{cs}$.
 Noteworthy that the \ha\ model is not sensitive to the parameter $\xi$. 
We adopt $\xi = 1.2$ that corresponds to the self-similar solution for 
  $q = 7$ and the uniform CS medium (Chevalier 1982).
The forward shock layer is assumed to be uniform with the density of $4\rho_0$. 
The high number density in the forward shock  ($n \sim 3 \times10^{10}$\cmq) implies 
  the fast electron-proton thermal equilibration compared to the expansion time, so 
  the electron temperature is $T_e = 1.6\times10^9v_{s,4}^2$, where $v_{s,4} = v_s/10^4$\kms.
The radiative cooling time is significantly longer than the expansion time, 
 however the Compton cooling time $t_{\sc C} =   
  1.2\times10^4r_{14}L_{43}^{-1}$ s can be comparable to the 
  expansion time; the postshock electrons can therefore cool by a factor of $\sim 2$. 
We therefore adopt $T_e = 10^9$\,K through the postshock layer.
Noteworthy that 
 the \ha\ profile is not sensitive to the variation of the electron temperature 
 even by a factor of ten.

The powerful early SN radiation results in the significant acceleration of the CS gas 
 with the velocity being maximal just before the shock and decreasing outwards. 
We describe the velocity distribution of the CS gas at the fixed moment by the expression 
\begin{equation}
 v(r) = (v_{ps} - v_{cs})\left(\frac{R_{cs} - r}{R_{cs} - R_s}\right)^s + v_{cs}\,,
 \end{equation}
   where $v_{ps}$ is the preshock CS velocity at $r = R_s$ 
   and $v_{cs}$ is the velocity of the undisturbed CS gas at $r = R_{cs}$.
 The value of power index is $s \approx 1.6$ for optimal  \ha\ models.  
 
 \vspace{1cm}
 
\section{Results}

\subsection{CS gas velocity}

The radiation transfer of  \ha\ photons in the CS shell is treated using the Monte Carlo technique.
The  \ha\ emission is dominated by the recombination mechanism, so the emissivity in the uniform 
  CS envelope is assumed to be constant along the radius. 
The model radiation transfer takes into account resonanant scattering in \ha\ 
   in the Sobolev approximation.
  However the previous modelling of the \ha\ at 10.3\,h (Chugai 2020a) indicates that 
   the Sobolev optical depth in \ha\ should be negligibly small.
 This situation reflects strong depopulation of the hydrogen second level due 
   to the high photoionization rate by the SN radiation. 
   
The angle-averaged frequency redistribution function (Mihalas 1978) is applied for the photon 
   scattering on thermal electrons.
The \ha\ profile is not sensitive to the exact value of 
   the CS electron temperature since the broadenning is dominanted by the high velocities of the 
   radiatively accelerated CS gas. 
Nevertheless the modelling takes into account the evolution of the electron temperature. 
At the first iteration we use the value $T_e = 4\times10^4$\,K for 
   all the considered epochs.
With this temperature we recover CS velocity from \ha, the bolometric luminosity and 
 the effective temperature.
This temperature is adopted as the electron temperature  and 
  these values of $T_e$ are used for the final \ha\ models.  
The radiation transfer includes a diffuse reflection of photons from the photosphere. 
However this effect for \ha\ photons is equivalent to zero albedo because reflected 
   photons scatter in the far blue wing due to 
  the high photosphere velocity of $\gtrsim 26000$\kms\ at the considered epoch (Fig.~\ref{fig:dyn}).

\begin{figure}
    \epsfxsize=19cm
	\hspace{-2cm}\epsffile{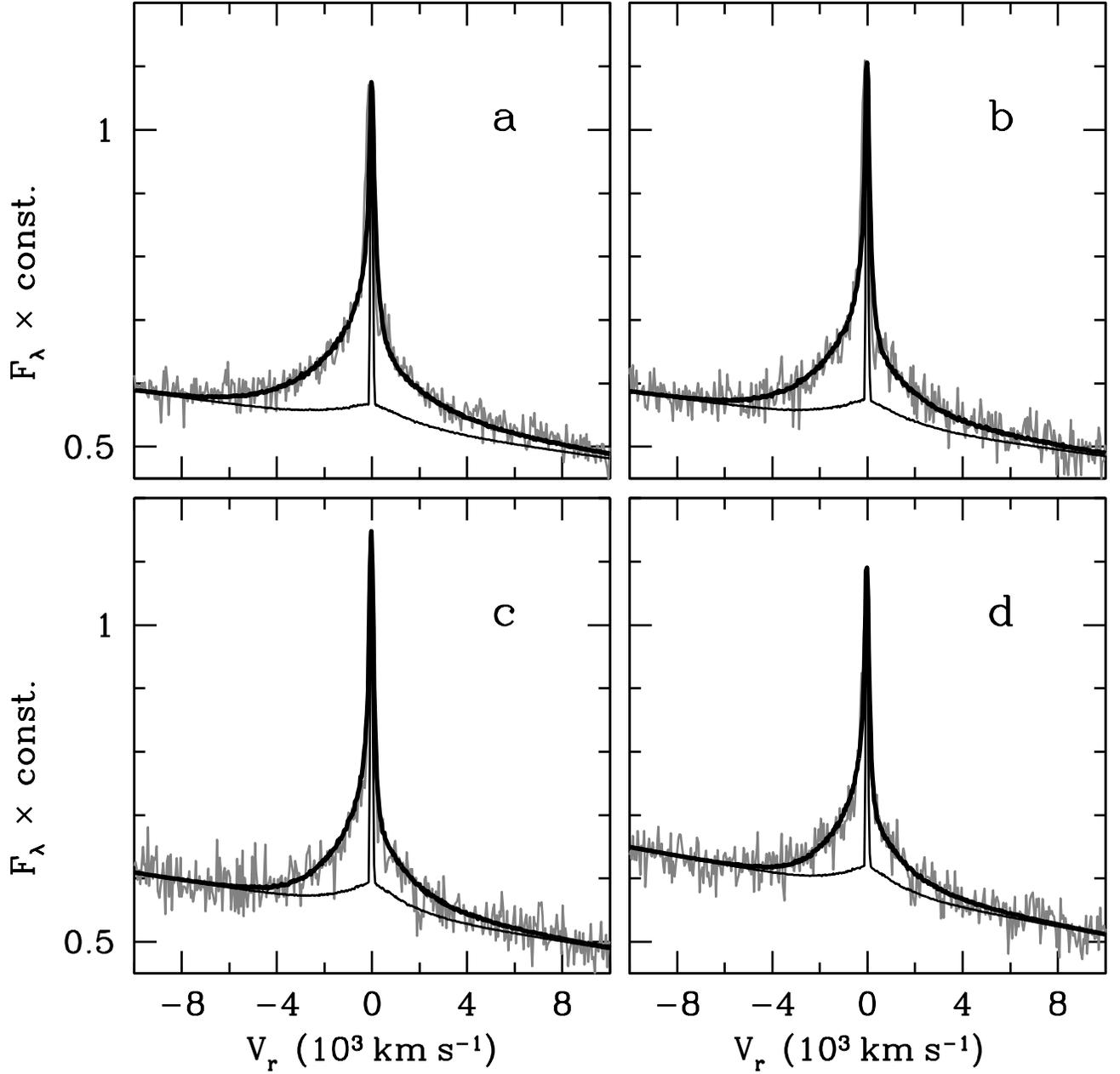}
	\caption{\ha\ emission in spectra of SN~2013fs.
	Model \ha\ ({\em thick} line) is overploted on observed spectra ({\em grey}) taken
	at the moments 0.26\,d (panel {\bf a}), 0.3\,d ({\bf b}), 0.37\,d ({\bf c}), and 0.42\,d 
	({\bf d}) after the shock breakout.
    {\em Thin black} line shows models with the constant CS gas velocity of 100\kms.
    }
	\label{fig:sp} 
\end{figure}

The optimal \ha\ models are overplotted on the observed low resolution 
  (160\kms) spectra  between 6\,h and 10\,h  (Fig. \ref{fig:sp}) with parameters 
   given in the Table 1. 
The table columns contain the time after the shock breakout, the CDS radius, the electron temperature of 
  the CS gas, the preshock Thomson optical depth, and the preshock velocity inferred from the 
  \ha\ fit. 
The bottom line includes the result obtained earlier for the high resolution spectrum 
  at 10.3\,h (Chugai 2020a).
The uncertainty of the inferred velocity is in the range of 20\%. 
The similar uncertainty refers to the Thomson optical depth recovered from \ha. 
The primary indicator of the Thomson optical depth is the profile skewing towards blue being 
  apparent in all the profiles of Fig.~\ref{fig:sp}.
However, $\tau_{\sc T}$ in Table 1 are taken from the 
  interaction model for consistency; these values are used for the \ha\ model.
  
The crucial role of the radiative acceleration of the CS gas for the \ha\ profile is 
   emphasised by the \ha\ computed without radiative acceleration effect, and using 
   the constant velocity of 100\kms\ for the CS shell (Fig. \ref{fig:sp}).
It is clear that the Thomson scattering exclusively cannot account for the observed \ha.   
This differs SN~2013fs from SN~1998S where the Thomson scattering dominates over the 
  effect of the moderate radiative acceleration of 1000\kms\ (Chugai 2001).
 
\subsection{Early bolometric luminosity}
\label{sec:bol}

Preshock CS velocity $v_{ps}$ is a direct indicator of the radiation energy $E_r$ emitted 
  by the supernova between the shock breakout and the observation epoch.
The radiative force is dominated by the Thomson scattering 
 for conditions in the CS shell (cf. Chugai et al. 2002).
Neglecting the CS gas displacement the solution of the equation of motion of the CS gas 
  in the field of SN radiation results in the preshock velocity 
 \begin{equation}
   v_{ps} = \frac{k_{\sc T}E_r}{4\pi R_s^2c}\,, 
 \label{eq:accel}  
 \end{equation}  
   where $k_{\sc T} = 0.34$\cmsqg\ is the coefficient of the Thomson scattering,
   $c$ is the speed of light.
   
\begin{figure}
	 \epsfxsize=19cm
	\hspace{-2cm}\epsffile{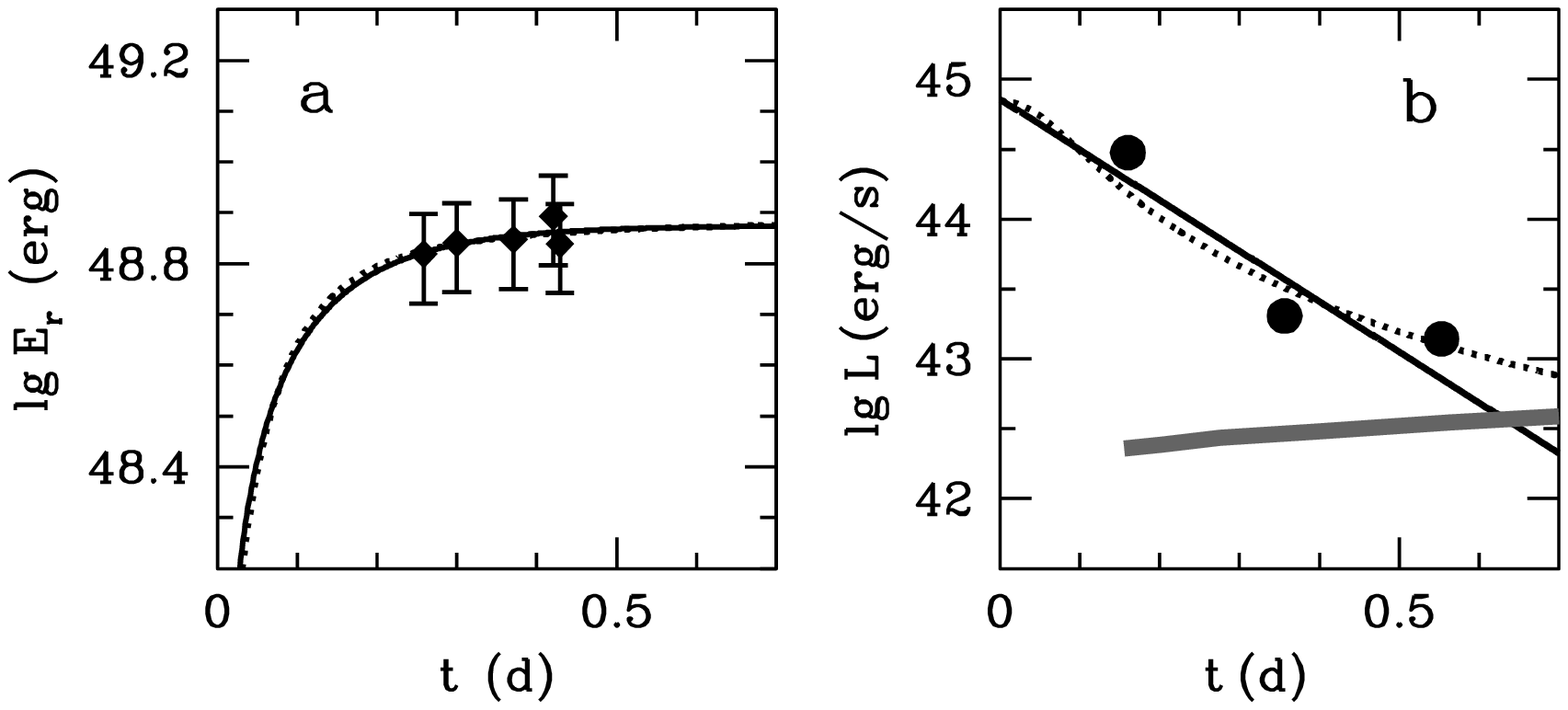}
	\caption{Radiation energy emitted by SN~2013fs (panel {\bf a}) recovered from the 
	preshock velocities ({\em diamonds}) with the overploted luminosity presented by the exponential 
	luminosity decline ({\em solid } line) and the power law ({\em dotted}). 
	Panel {\bf b} shows the recovered luminosity for the exponential law ({\em black} line) 
	and power law ({\em dotted} line)
	that fit the radiation energy in panel {\bf a}. 
	{\em Circles } correspond to the reported luminosity (Yaron et al. 2017)
	recovered from the multiband photometry based on the black body approximation, while the {\em grey} line is the luminosity recovered from the multiband photometry based on the SED reconstruction. 
    }
	\label{fig:bol}
\end{figure}

 The inferred $v_{ps}$ values (Table 1) with the equation (\ref{eq:accel}) permit us to 
    recover $E_r$ for the explored  moments.
 We apply two different descriptions of the initial stage of the luminosity decline: 
   the exponential law  $L = L_0\exp{(-t/t_0)}$ and the power law $L = L_0/[1 + (t/t_0)^p]$.
 In each case parameters are determined by the $\chi^2$  minimization.   
For the exponential law $t_0 = 0.12$\,d and  $L_0 = 7.23\times 10^{44}$\ergs, whereas 
  for the power law $t_0 = 0.12$\,d, $L_0 = 5.8\times 10^{44}$\ergs, and  $p = 2.6$ 
  (Fig. \ref{fig:bol}).
Both descriptions for the luminosity coincide within 10\%, while the energy radiated 
  during initial 0.5\,d in both cases is $7.4\times10^{48}$\,erg.
The relative error of $E_r$ is the same as that of the velocity, i.e. 20\%.  
 
 The recovered bolometric luminosity is compared (Fig. \ref{fig:bol}, panel b) to the      
    reported bolometric 
   light curves obtained from the multiband photometry in two ways (Yaron et al. 2017).
 The first approach suggests the reconstruction of the spectral energy distribution (SED), 
   while the second method is based on the recovered blackbody  
  temperature, photospheric radius, and the blackbody luminosity  $L = 4\pi R^2\sigma T^4$. 
 These two reported versions differ by a factor of 100 (Fig. \ref{fig:bol}, panel b) at the first   
    observational   
    epoch that emphasises difficulties of the early light curve reconstruction 
     from the photometric data.
 Amazingly, our light curve is consistent with the blackbody version of the reported light  
    curve despite of radically different methods.
    
 It is noteworthy that the early bolometric light curve of SN~2013fs recovered 
   from the radiative acceleration effect in \ha\ does not requires  
   the photometry, the distance, and the extinction.
 The point is that this method exploits only measurements of the velocities based 
  on spectra expressed in relative fluxes.

\section{Conclusion}
 
The paper has been aimed at the reconstruction of the early bolometric light curve 
  of SN~2013fs exploiting effect of the radiative acceleration of the CS gas being 
 apparent in the \ha\ emission.
The proposed method turns out to be a success in the case of SN~2013fs 
  thanks to the unique set of Keck-I spectra between 6 -- 10.3\,h after the shock breakout. 
The attractive feature of the novel method is that the early bolometric light curve of SN~2013fs 
  is recovered disregarding the photometry, distance, and extinction.

  Some uncertainty might stem from the choice of a function that approximates  
    the luminosity decline after the shock breackout.
 In fact this arbitrariness does not affect the result.
 Two different choices, exponential and power law, result in the bolometric light curves  
  that agree with each other within 10\%; moreover the total radiation emitted during 
  0.5 day after the shock breakout in both cases is the same.
Suprisingly, that the recovered bolometric luminosity is close to the luminosity 
  estimated using blackbody approximation for the moments 0.16, 0.36, and 0.55\,d 
  ((Yaron et al. 2017).
The agreement between results obtained by two completely different approaches indicates 
  that both methods catch behavior of the real luminosity of SN~2013fs at the initial stage.
  
Yet unlike the case of SN 2013fs with the known distance 47-51 Mpc (NED) and small extinction 
   (Yaron et al. 2017), for nearby SNe~IIP  with less certain distance and extinction 
   sytematic errors in the bolometric luminosity recovered from the photometry can be 
   large.
All three factors of the uncertainty (distance, extinction, and photometric incompleteness) 
  are lacking in the new method. 
It should be emphasised potential significance of the new method for the reconstruction 
  of the early bolometric luminosity of a future SN~IIP in our Galaxy, since in that case 
  the distance and the extinction will be known likely with large uncertainties.
Obviously, a successfull application of the proposed method requires the 
   observation of a set of spectra during the first day after the explosion.


\pagebreak   

\pagebreak

\begin{table}[t]   
	
	\vspace{6mm}
	\centering
	{{\em Table 1.} Parametrs of \ha\ model}
	\label{tab:spar} 
	
	\vspace{5mm}\begin{tabular}{l|c|c|c|c} 
		\hline
 Day   &  $R_{cds}$ ($10^{14}$\,cm) & $T_e$   & $\tau_{\sc T}$ & $v_{ps}$ (\kms)  \\
\hline   
0.258 &   0.74     &  60000$\pm3000$ &      1.9    &  7500$\pm1500$  \\
0.30  &   0.85     &  50000$\pm2500$ &      1.9    &  6000$\pm1200$   \\
0.371 &   1.05     &  40000$\pm2000$  &     1.8    &  4000$\pm800$  \\
0.421 &   1.18     &  35000$\pm1800$  &     1.7    &  3500$\pm700$ \\
0.423 &   1.19     &  35000$\pm1800$ &     1.7    &  3000$\pm600$  \\
\hline

	\end{tabular}
\end{table}

\begin{thebibliography}
\bibitem{}
Badnell N.R., Bautista M.A., Butler K., et al., Mon. Not. R. Astron. Soc. {\bf 360}, 458 (2005)
\bibitem{}
Chevalier R.A.,  Astrophys. J. {\bf 259}, 302 (1982)
\bibitem{}
Chevalier R.A., Fransson C., Nymark T.K.,  Astrophys. J. {\bf 641}, 1029 (2006)
\bibitem{}
Chevalier R.A., Fundamentals of Cosmic Physics. {\bf 7}, 1 (1981)
\bibitem{}
Chugai N.N., Astron. Lett. {\bf 46}, 319 (2020a)
\bibitem{}
Chugai N.N., Mon. Not. R. Astron. Soc. {\bf 494}, L86 (2020b)
\bibitem{}
Chugai N.N., Blinnikov S.I., Fassia A. et al. Mon. Not. R. Astron. Soc. {\bf 330},
  473 (2002)
\bibitem{}
Chugai N.N., Mon. Not. R. Astron. Soc. {\bf 326}, 1448 (2001)
\bibitem{}  
 Dessart L., Hillier D.J., Audit E., Astron. Astrophys. {\bf 605}, A83 (2017)
\bibitem{} 
 Grasberg E.K., Imshennik V.S., Nadyozhin D.K., Astrophys. Space Sci. {\bf 10}, 3 (1971)
\bibitem{}
 Groh J.H., Astron. Astrophys. {\bf 572}, L11 (2014)
\bibitem{}
Khazov D., Yaron O., Gal-Yam A. et al., Astrophys. J. {\bf 818}, 3 (2016)
\bibitem{}
Quimby R.M, Wheeler J.C., H{\"o}flich P. et al., Astrophys. J. {\bf 666}, 1093 (2007)
\bibitem{} 
Shivvers I., Groh J.H., Mauerhan J. C. et al., Astrophys. J. {\bf 806}, 213 (2015)
\bibitem{} 
Utrobin V.P, Chugai N.N.,  Astron. Astrophys. {\bf 555}, A145 (2013)
\bibitem{}
Yaron O., Perley D.A., Gal-Yam A. et al., Nature Physics {\bf 13}, 510 (2017)
\bibitem{}
Yaron O., Gal-Yam A., Publ. Astron. Soc. Pacific {\bf 124}, 668 (2012)


\end{thebibliography}
\end{document}